\documentstyle[emulateapj,onecolfloat,psfig]{article}   

\begin{document}
\twocolumn[      

\title{Occultation Searches for Kuiper Belt Objects}

\author{Asantha Cooray and Alison J. Farmer}
\affil{Theoretical Astrophysics, Mail Code 130-33, California Institute of Technology, Pasadena, CA 91125\\
E-mail:(asante,ajf)@tapir.caltech.edu}

%------------------------------------------------------------------------------

\begin{abstract}
The occultation of background stellar sources by foreground Kuiper Belt Objects (KBOs) can be used to 
survey physical properties of the KBO population. We discuss statistics related to a KBO occultation survey, 
such as the event duration
distribution, and suggest that occultation searches can be effectively used to probe the 
KBO size distribution below 10 km. In particular, we suggest that occultation surveys may be best 
suited to search for a turnover radius in the KBO size distribution due to collisions 
between small-size objects. For occultation surveys
that monitor stellar sources near the ecliptic over a few square 
degrees, with time sampling intervals of order 0.1 sec and sensitivity to flux variations of a few percent or more, 
a turnover radius between 0.1 and 1.0 km can be probed. 
While occultation surveys will probe the low-radius limit and imaging surveys 
will detect KBOs of size 100 km or more, statistics of objects 
with sizes in the intermediate range of around 1 km to 100 km will likely remain 
unattainable.
\end{abstract}
%------------------------------------------------------------------------------
% User-supplied List of keywords.

]

%------------------------------------------------------------------------------
\section{Introduction}

The presence of minor bodies beyond the orbit of Neptune is now well established; current estimates suggest that
there is a total of order $10^5$ Kuiper Belt Objects (KBOs) with sizes greater than 100 km (see Luu
\& Jewitt 2002 for a recent review). 
Extending KBO population statistics to lower sizes than currently known
is essential to understanding the formation and evolution
of these minor planets.  Since a kilometer-sized KBO is expected to have a magnitude $R\sim$ 30,
it is unlikely that imaging surveys will be able to obtain reliable statistics of
small-size objects in the Kuiper Belt.

The occultation of background stellar sources by foreground objects has been well utilized as a probe of
foreground source properties (see Elliot \& Olkin 1996 for a review). In fact, suggestions have already been made
to detect KBOs via occultation observations (Bailey 1976; Brown \&
Webster 1997; Roques \& Moncuquet 2000). On the observational side, the Taiwanese-American Occultation Survey 
(TAOS; Liang et al. 2002) plans to conduct an occultation search using an array of three to four
robotic telescopes simultaneously monitoring the same 3 square degrees of sky.

In this Letter, we discuss statistics related to KBO occultations and physical properties of the KBO
population that can be probed with a wide-field 
survey targeting the ecliptic area.
For reasonable parameters that can be achieved with ground-based telescopes,
we suggest that 
occultation observations are effective at finding small-size KBOs with radii between $\sim$ 0.1 and 10 km.
At this low end of the KBO size distribution, one no longer expects the power-law size distribution
observed for KBOs larger than 100 km to hold. The frequent collisions between small-size objects, especially
at early times during the formation of the large bodies, lead to their destruction to dust such that a turnover radius 
is expected at some characteristic size scale (Kenyon 2002). 
For typical values corresponding to the Kuiper Belt, with a surface
density, $\Sigma\sim 10^{-4}$ g cm$^{-2}$, an angular velocity,
$\Omega \sim 2 \pi/250$ year$^{-1}$, a KBO density, $\rho\sim 1$ g cm$^{-3}$, and an age for the belt, $\tau\sim 4 \times 10^9$ years, an estimate for the 
turnover radius is $R_c \sim \tau \Sigma \Omega/\rho \sim 0.1$ km.
We suggest that occultation surveys have the potential to either detect or constrain
an expected turnover radius, which will, in return, help in the 
theoretical understanding of the formation and evolution of the Kuiper Belt.

The discussion is organized as follows. In the next Section, 
we will summarize important ingredients necessary for a calculation of statistics related to KBO occultations. 
In addition to the  size distribution of KBOs, we improve previous estimates on the KBO occultation rate by paying 
particular  attention to the size distribution of background stars projected at KBO distances. As we will discuss later, 
the occultation duration distribution is determined by both
projected star sizes and KBO sizes. In \S~3, we will present our results and conclude  with a summary.

\section{Calculational Ingredients}

One can characterize an occultation 
of a background star by a foreground KBO with two parameters: the 
duration $\Delta T$ of the occultation  and the fractional flux
decrease $A$ of the background star during the occultation. 
In terms of physical parameters, these two observables are given by 
\begin{equation}
\Delta T = \frac{2(R+R_\star)}{v}, \; {\rm and} \;
A=\left(\frac{R}{R_\star}\right)^2,
\end{equation}
respectively. Here, $v$ is the relative velocity of the KBO with
respect to the observer while  $R$ and $R_\star$ are respectively the radii of the
KBO object and the background star; the latter is the
projected stellar radius at the distance of the KBO. We have assumed
no flux contribution from the KBO.
In writing the duration of the event, we have taken into account the 
finite size of stellar sources beyond the point-source approximation
(cf. Brown \& Webster 1997); we shall see that this is important.

\subsection{KBO distribution}

We make use of the form for the differential size 
distribution of KBOs (suggested by Kenyon 2002) as a function of radius $R$ such that
\begin{equation}
\frac{d N_{\rm{kbo}}}{dR} = \left\{ \begin{array}{ll} A_0 \left(\frac{R}{R_c}\right)^{-q_1}\;  \quad \quad \quad \quad \quad R > R_c \, \\
A_0 \left(\frac{R}{R_c}\right)^{-q_2}\;  \quad \quad \quad \quad \quad R \leq R_c \, ,
\end{array} \right.
\label{eqn:kbosize}
\end{equation}
where $R_c$  is the turnover radius, which is expected to be in the range 0.1 to 10 km.
The power-law slope, $q_1$, has been estimated to be $\sim$ 4 when $R > 100$ km (Trujillo et al. 2001)
and we will set $q_1=4$ hereafter. For the present discussion, 
we treat $q_2$ and $R_c$ as parameters that can potentially be probed with occultation 
observations. The normalization constant $A_0$ is set by requiring that the total number  
of KBOs above 100 km agrees with the estimated number of classical and scattered KBOs,
$7 \times 10^4$, based on observations so far (Trujillo et al. 2001).

The differential distribution of KBOs with distance $D$, $df/dD$, 
is taken to scale as $D^{-\gamma}$, with minimum and maximum
distances 30 AU and 50 AU respectively (Brown \& Webster 1997; Kenyon 2002). 
We set $\gamma=2$; our results are not strongly sensitive to the choice of $\gamma$. The
KBOs are taken to be uniformly distributed in azimuth, with a
Gaussian distribution $P(i)$ in inclination angle $i$ from the
ecliptic. Following measurements by Trujillo
et al. (2001), we take the standard deviation of this Gaussian to be $15^\circ$.

\begin{figure}[t]
\centerline{\psfig{file=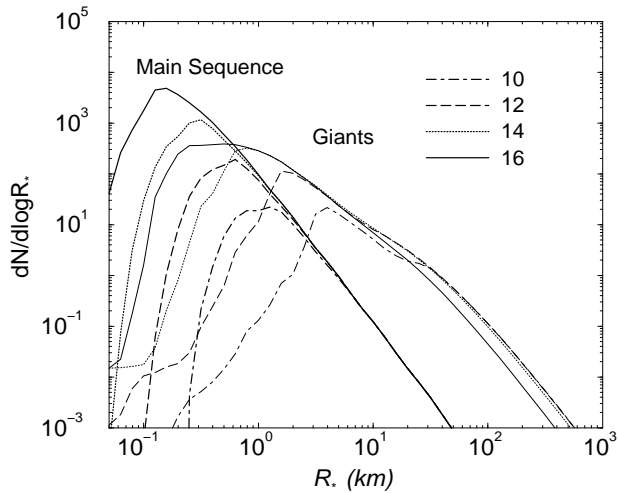,width=3.2in,angle=-90}}
\caption{The distribution of stellar radii projected at a distance of 40 AU. We show the distribution as a function of
magnitude limit in the V-band as labeled on the figure.
The left set of four curves corresponds to main
sequence stars while the right set is for giants. The counts are normalized
to a total of $\sim$ 56,000 stars (down to $V\sim 17-18$) 
in the three square degree ``crowded field'' centered on $l=170^\circ, b=-9^\circ$ that is planned to be monitored by the
TAOS survey.}
\label{fig:lc}
\end{figure}

\subsection{Stellar distribution}

\begin{figure}[t]
\centerline{\psfig{file=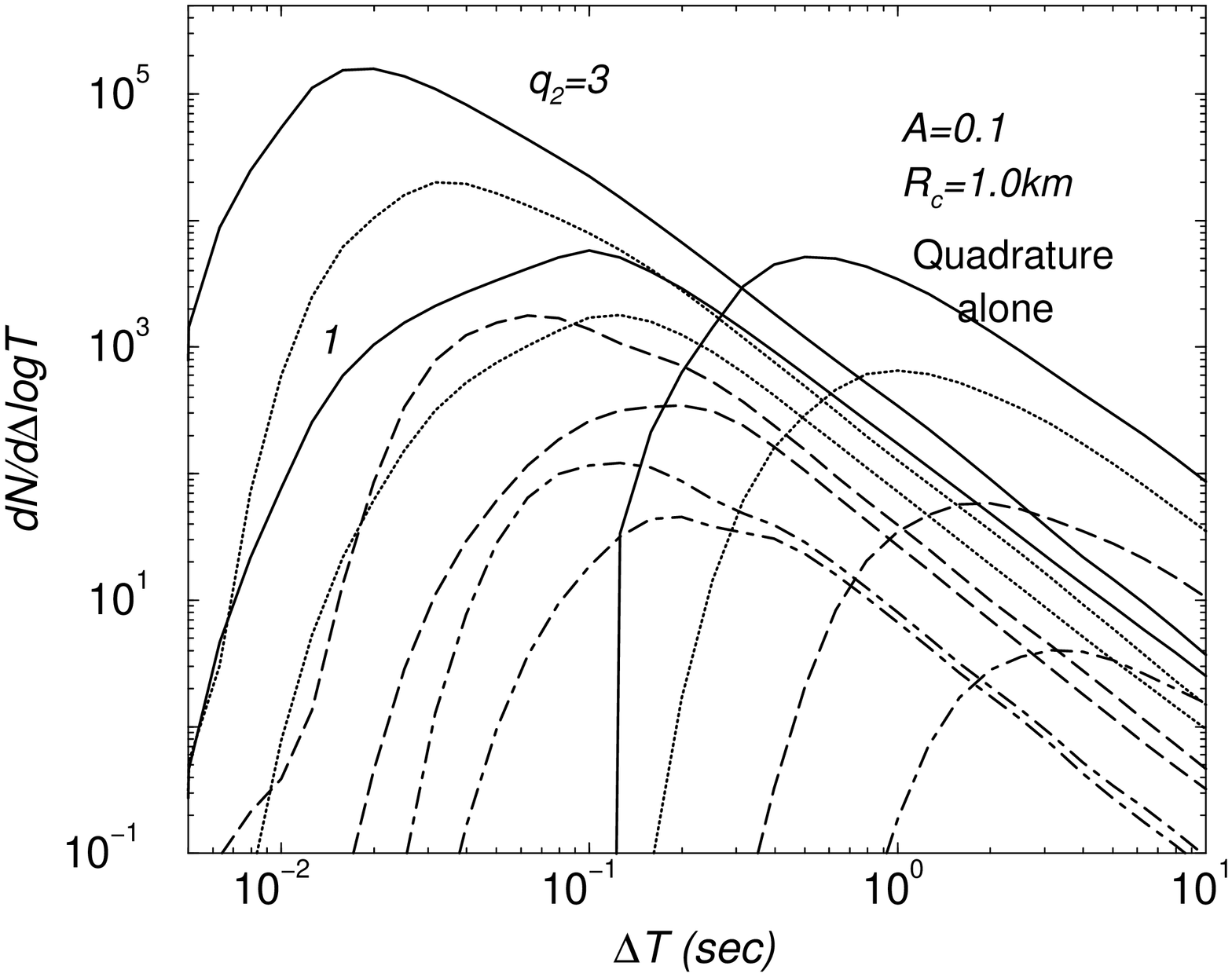,width=3.2in,angle=-90}}
\centerline{\psfig{file=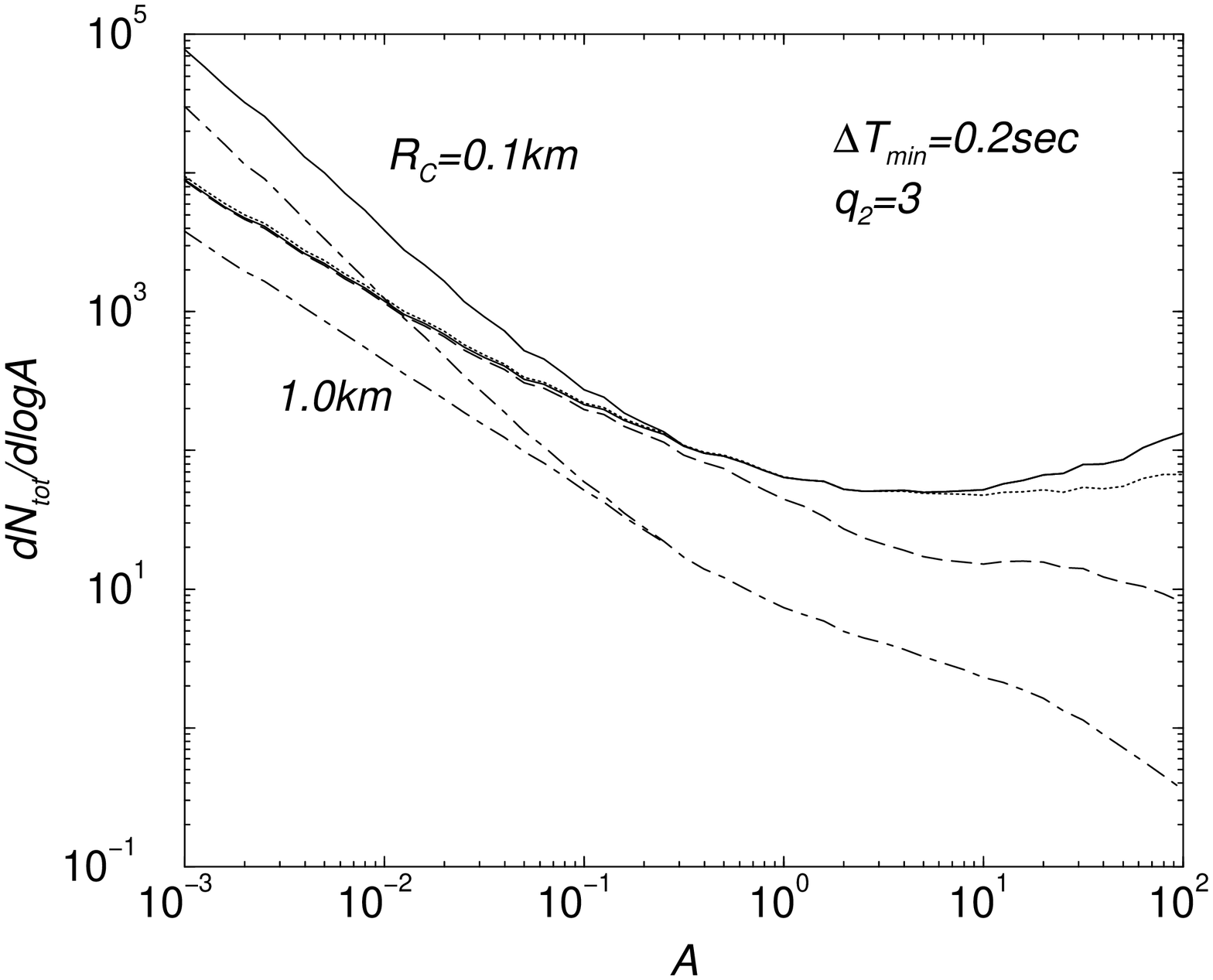,width=3.2in,angle=-90}}
\caption{The KBO occultation duration (top) and flux change (bottom) distributions, with parameters for the
KBO size distribution as shown, for a one-year survey of the same three square
degrees of sky. The linestyles denote magnitude limits as in Fig.~1.
In the top plot, we show the variation in the duration distribution as a function of $q_2$, 
the slope of the KBO radius distribution below the turnover radius, assuming $R_c=1$ km. For comparison,
we also show the duration distribution for a year long survey that targets quadrature; while quadrature gives
an increase in duration, this comes at the expense of a reduction in
the total number of events by the same factor.
In the bottom plot, we show variations to the flux change distribution as a function of $R_c$ assuming $q_2=3$.}
\label{fig:deltaT}
\end{figure}

Following standard approaches discussed in the literature, 
we construct a representative stellar angular radius distribution by
first creating a population of single stars born at a constant rate
over the past 10 Gyr, according to the initial mass function of Kroupa,
Tout \& Gilmore (1993). This was then evolved to the present day using the
\textsc{bse} stellar evolution code (Hurley, Pols \& Tout 2001), which
provides absolute magnitudes and stellar radii as a function of time. Finally, the
stars were distributed according to a simple exponential Galactic disk
model, with disk scale radius 3.5 kpc and scale height 200 pc. 
Incorporating extinction corrections from Bahcall \& Soneira (1980),
we selected magnitude-limited samples to produce differential stellar
angular size distributions, $dN/d \log R_\star$,  
for a given field, where $R_\star$ is the projected stellar radius at
the distance of the Kuiper Belt. 

The calculated angular size distribution (Fig.~\ref{fig:lc}) rises
towards small projected sizes, up to a cutoff related to the
maximum distance to which stars are seen in a given sample. The
position of the cutoff hence shifts to smaller sizes as the magnitude limit
is increased. Beyond $V\sim 12$, where the finite extent of the Galaxy
becomes important, decreasing the Galactic latitude $|b|$ of the field
has the same effect. The peak at small angular sizes is composed of
comparable numbers of A, F, and G main sequence stars.
The distribution can be observationally
obtained for a given field, based on colors and magnitudes (Van Belle 1999) and is not
likely to be major source of uncertainty. We have however neglected the presence of unresolved binary companions to half of these
stars (Duquennoy \& Mayor 1991). 
Due to the neglect of light from unresolved companions,
we may have overestimated the number of events above a given amplification limit.
The neglect of the extra cross section presented by companions,
however, underestimates the event rate.
Since in most binaries one expects the
primary to be considerably brighter than the secondary, it is likely that the resulting effect is small.
We expect our predictions for the stellar size distribution
to be accurate to within a factor of 2.

\subsection{Relative Velocity}

Relative to an Earth-based observer, the velocity of a KBO  at
distance $D$ and observational angle $\phi$ is
\begin{equation}
v_{\rm KBO}(D,\phi) \approx v_{\earth}\left[\cos\phi-\sqrt{\frac{D_{\earth}}{D}}\right] \, .
\end{equation}
Here, $v_{\earth}=30$ km sec$^{-1}$ and $D_{\earth}=1$ AU. As $\phi$ is varied, $v_{\rm KBO}(D,\phi)$  
ranges from 0 at quadrature to $30(1-\sqrt{D_{\earth}/D})$ km sec$^{-1}$ at opposition. 
For a KBO at a distance of 40 AU, the maximum velocity is $\sim$ 25 km sec$^{-1}$.
Since we will concentrate on 
KBO occultation surveys that target areas around the ecliptic, where the KBO density
is highest, we do not include the dependence of KBO relative velocity
on inclination angle. For surveys that monitor specific fields over long periods, such as TAOS, note that $\phi$ changes during the year. 
An alternative possibility is to monitor the night sky while keeping the position angle constant. 

\subsection{Diffraction}

An additional consideration with respect to KBO occultations involves diffraction;
the Fresnel scale, $R_{\rm F}=\sqrt{\lambda D/2 \pi}$, at a distance of 40 AU, for $\lambda=500$ nm, 
is $\sim$ 0.7 km and is typical of both KBO and projected stellar sizes.
The resulting effect due to diffraction can be described as an increase in the duration of the occultation
with a corresponding decrease in the flux decrement.
We therefore modify Eq.~(1), by replacing $(R+R_\star)$ with $R_{\rm diff}$, where
$R_{\rm diff}$ is the effective occultation radius in the presence of diffraction fringes.
We model this following Roques \& Moncuquet (2002), and assuming that KBOs are regularly shaped objects.
Note that finite background star sizes, especially when $R_\star > R_{\rm F}$, reduce effects due to 
diffraction such that a geometrical description is valid with $R_{\rm diff} \rightarrow R+R_\star$
for long duration events, at the level of 0.1 seconds or more. We will highlight the gain involved with diffraction,
especially for surveys that can reach sensitivities to flux variations at the level of 1\% or below.

\section{Occultation statistics}

With the distributions of $R$, $R_\star$, and $v$ established above, we can now write the
expected number of KBO occultations during a total 
 observing time $T_{\rm tot}$ as
\begin{eqnarray}
&&N_{\rm tot} = \int d\phi \int di P(i) \int \frac{dD}{D^2}
  \frac{df}{dD} \int dR_\star \frac{dN}{dR_\star} \nonumber \\
  &\times&\int dR \frac{d N_{\rm KBO}}{dR} 2R_{\rm diff} v_{\rm KBO}(D,\phi) T_{\rm tot}\, .
\label{eqn:N}
\end{eqnarray}
The integrals are taken such that one allows for a minimum fractional dimming, $A_{\rm min}$,
and a minimum sampling interval in time, $\Delta T_{\rm min}$, with which flux from
stars down to some magnitude limit can be monitored. In a real
survey, $A_{\rm min}$ will be a function of magnitude, and we consider
an example of this later in the Section.
Instead of discussing the total number of occultations, we consider the fractional distribution 
of occultations in terms of observables, mainly event duration, fractional flux variation, and KBO radius.
These are calculated by introducing the necessary Dirac delta function
before taking integrals in Eq.~\ref{eqn:N}, such that, say, $dN_{\rm tot}/d\delta T = \int d\phi ... \int dR\; dN_{\rm KB0}/dR
2 R_{\rm diff} v T_{\rm tot} \delta_D(\delta T - 2 R_{\rm diff}/v)$.

In Fig.~2, we show the expected number of KBO occultations as a
function of duration, for a year-long survey of a given field, using  
our geometrical description for occultations, with no corrections due to diffraction. 
Most of the occultation events are of short duration, and are
associated with the main sequence stars producing the peak at small
angular size. For comparison, we also
show the duration distribution towards quadrature.
While durations increase by an order of magnitude or more,
there is an equivalent reduction in the total number of events,  
which can be understood by noting that the duration scales as $v^{-1}$ while
the total number of occultations scales as $v$. For a long term survey
of a given field, 
the distribution predicted towards opposition alone 
is similar to that obtained when one integrates over position angle
throughout the year. This is due to the fact that,
away from quadrature, which is highly localized, all other position angles lead to comparable velocities.

The predictions in Fig.~2 reveal that for
surveys which are limited to large timing intervals,
of order 0.1 sec or more, but fainter magnitudes, V $\sim$ 16, 
observations toward quadrature are desirable; this, however, requires
monitoring of stars in a number of different fields, since the background sky
shifts with respect to the direction of quadrature during a long period of observation.
With small improvements in the minimum duration to which a survey is
sensitive, one can obtain a significant increase in event rate by
observing, say, one field throughout the year, instead of targeting
the low-velocity quadrature region. The use of a single field also
allows better control of systematics and a reduction of confusions,
such as those due to intrinsic stellar variabilities.

The bottom plot of Fig.~2 shows the expected distribution of flux variations. 
This distribution has a power-law behavior when $A < 0.1$ and flattens out as $A$ approaches one. 
Note that, in a strict sense, the maximum value of $A$ is one, when a KBO fully occults the
background star. Here, for illustration purposes, 
we have allowed $A$ to vary beyond one in order to show the flattening of this distribution.

\begin{figure}[t]
\centerline{\psfig{file=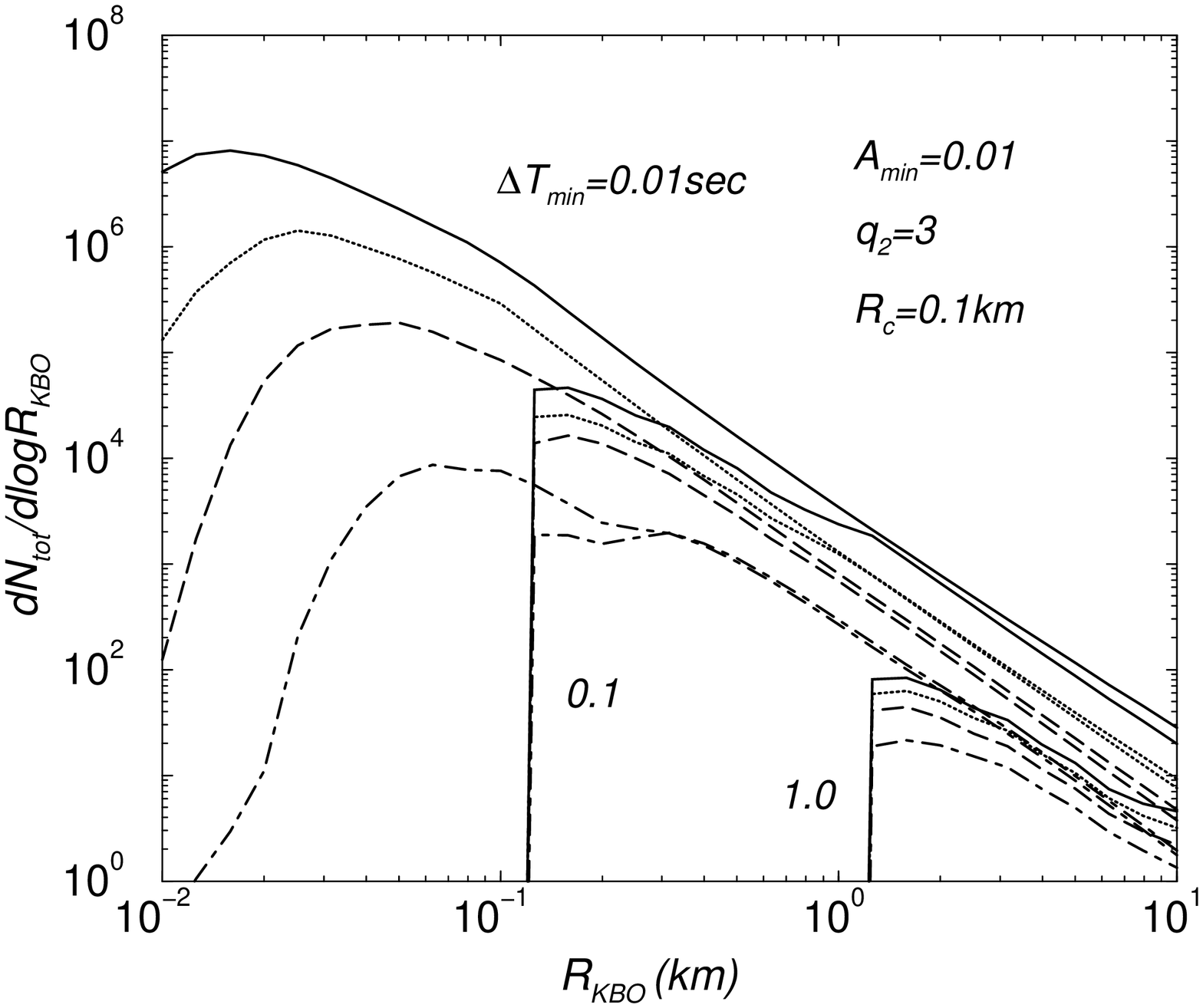,width=3.2in,angle=-90}}
\centerline{\psfig{file=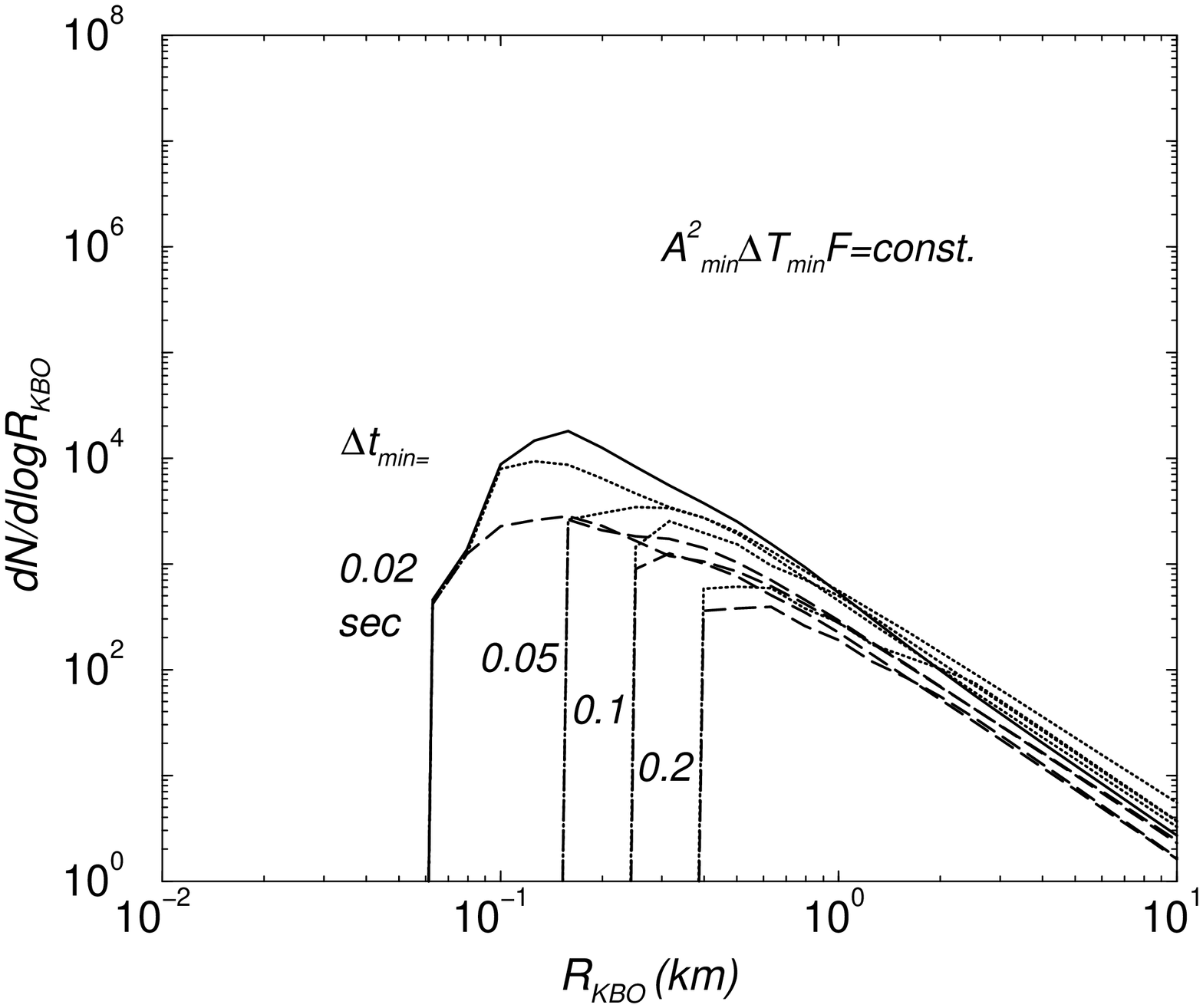,width=3.2in,angle=-90}}
\caption{The KBO occultation distribution as a function of the KBO size.
For occultation surveys with  sensitivity to flux variations of order 1\%
and sampling intervals of order 100 msec, KBOs down to a radius
of $\sim$ 0.1 km can be probed.
For such surveys, one does not gain significantly by monitoring stars with
magnitudes fainter than
$\sim$ 14. In the top plot, we also show the expected boost in number of
events due to diffraction for the best case scenario involving a sensitivity to flux variations at a
level of 1\%. In the bottom plot, we consider a
single experiment for which $F A_{\rm min}^2 \Delta T_{\rm min}$=constant 
with normalization set at $A_{\rm min}=10$\% at 14th magnitude with
$\Delta T_{\rm min}=0.2$ sec. 
We improve $\Delta T_{\rm min}$ as shown on the plot, for a survey with $V<14$, except for the case of $\Delta
T_{\rm min}=0.02$ sec, for which we show curves for limiting magnitudes of 12, 
14, and 16. Few additional events are gained by monitoring deeper than
$V=14$, especially when one considers the additional data reduction required.}
\label{fig:R}
\end{figure}

In Fig.~2, we also show variations in the
expected number of observed occultations induced by variations in 
the cutoff radius $R_c$, and the slope $q_2$ below the cutoff radius. 
With sufficient sensitivity to flux variations and with a high sampling rate,
one can also probe these two parameters; for example,
the expected number of occultations with flux variations between 1\% and
10\% differs by two orders of
magnitude when $R_c$ is varied between 0.1 km and 1 km.

In Fig.~3, we show the expected number of occultations as a function of KBO radius. For surveys with sampling intervals of 0.1 seconds and sensitivity to normalized flux decreases of 1\% or better, one probes the size distribution
down to a radius of 0.1 km.  The cutoff
corresponds to the KBO radius producing the smallest detectable flux
decrease $A_{\rm min}$ for a star whose size leads to an occultation with $\Delta T = \Delta
T_{\rm min}$.  We also highlight the gain associated with diffraction here.
With a detection threshold of order 1\% in flux, and a time sampling interval of 0.01 sec, diffraction
boosts the number of expected occultations by an order of magnitude. 
It is unlikely that  occultation surveys with ground based small telescopes can achieve the high flux sensitivity to
see effects related to diffraction, though
targeted surveys of smaller stars with 8m-class telescopes may be used to exploit the gain associated with
diffraction (Roques \& Moncuquet 2000).

We now consider a real experiment: the planned TAOS
survey (Liang et al. 2002). We take into account the presence of photon noise, which leads to a
decrease in flux sensitivity when individual exposures are shorter.
There is a tradeoff between sensitivity and sampling rate for a given experimental setup, which
can be expressed as $A^2_{\rm min} F \Delta T_{\rm
  min}$ = constant, where $F$ is stellar flux.
In Fig.~3 bottom plot, we show the expected distribution of occultations as a function of the KBO radius.
Here, we assume our fiducial KBO population ($q_1=4,q_2=3$ and $R_c=0.1$ km) and normalize the scaling relation 
with the expected TAOS values of $\Delta T_{\rm min}$ = 0.2 sec, and $A_{\rm min}$ =
10\% at 14th magnitude. As shown in the bottom plot of Fig.~3, with improvements to the sampling interval, and a
corresponding decrease in the flux sensitivity, one lowers the minimum KBO radius probed.
Following Fig~2, this is due to the rapid increase in the number of events with decreasing duration, when compared with
the distribution in terms of flux variations, which is slowly varying
when $A \sim$ 10\%.
We see that for surveys like TAOS, sensitive to
flux variations of order 10\% at a magnitude of 14 and time
intervals of a few tenths of a second, there is no significant gain in monitoring stars
fainter than 14th magnitude, since a large fraction of the 
detectable occultations are associated with bright stars of large
angular size, which have smaller mimimum
detectable flux variations and larger occultation durations, so more detectable occultations. The
additional data procesing involved with monitoring fainter stars could
also be cumbersome, given the diminishing returns.

To summarize, occultation surveys are ideal for probing a turnover radius in the size distribution
expected in certain theoretical models of formation and evolution of KBOs. To obtain adequate statistics, ground-based
surveys with smaller telescopes should monitor bright stars ($ V \la
14$) on and around the ecliptic, in fields where the density of stars
is also high.
In the case of the planned TAOS survey (Liang et al. 2002), it may be advisable to optimize the survey by
decreasing the minimum time interval with a corresponding decrease in sensitivity to flux variations.
Beyond the TAOS survey, we encourage
further occultation surveys to probe the low end of the size distribution since this is where theoretical 
models of KBOs can be tested in detail.

\smallskip
{\it Acknowledgments:} 
We thank R. Sari and P. Goldreich for useful discussions, C. Alcock for useful comments, and
J. Hurley for providing the \textsc{bse} code.
Partial support for this  research was supported by DOE 
and the Sherman Fairchild Foundation.

\end{document}